\documentclass[a4paper,12pt]{article}

\usepackage{amsmath,amssymb}
\usepackage{epsf}
\usepackage{cite}

\makeatletter

\def\half{{\frac{1}{2}}}
\def\thalf{{\tfrac{1}{2}}}
\def\p{\partial}

\def\unit{{1\kern-.65ex {\rm l}}}
\def\1{{1\kern-.65ex {\rm l}}}
\def\ap{{\alpha'}}

\def\gs{{g_{\rm s}}}

\def\sign{\mathop{\mathrm{sign}}\nolimits}

\def\Bh{{\widehat{B}}}

\def\Gh{{\widehat{G}}}

\def\Yh{{\widehat{Y}}}
\def\psih{{\widehat{\psi}}}

\def\Yt{{\widetilde{Y}}}

\def\phit{{\widetilde{\phi}}}
\def\psit{{\widetilde{\psi}}}

\def\cB{{\cal B}}
\def\cC{{\cal C}}

\def\cF{{\cal F}}

\def\cL{{\cal L}}

\def\cN{{\cal N}}

\def\bbR{{\mathbb{R}}}

\def\bbZ{{\mathbb{Z}}}

\newcount\hour \newcount\minute
\hour=\time \divide \hour by 60
\minute=\time
\def\now{%
\ifnum \hour<13
  \ifnum \hour=0 \advance \hour by 12 \number\hour:\else \number\hour:\fi%
     \ifnum \minute<10 0\fi%
     \number\minute%
\ A.M.%
\else \advance \hour by -12 \number\hour:%
  \ifnum \minute<10 0\fi%
  \number\minute%
  \ P.M.%
\fi%
}

\makeatother

\begin{document}

\baselineskip=18pt  
\numberwithin{equation}{section}  



%
\thispagestyle{empty}

\vspace*{-2cm} 
\begin{flushright}
YITP-13-66
\end{flushright}

\vspace*{2.5cm} 
\begin{center}
 {\LARGE Perturbative 3-charge microstate geometries\\[1ex] 
 in six dimensions}\\
 \vspace*{1.7cm}
 Masaki Shigemori\\
 \vspace*{1.0cm} 
Yukawa Institute for Theoretical Physics, Kyoto University\\
Sakyo-ku, Kyoto 606-8502, Japan\\
and\\
Hakubi Center, Kyoto University\\
Sakyo-ku, Kyoto 606-8501, Japan\\
 \vspace*{0.8cm} 
 {\tt sh{}i{}ge{}@yi{}t{{}}{{p}}.kyo{}t{}o{{-}}{{}}u.a{{c.}}j{}p} 
\end{center}
\vspace*{1.5cm}

\noindent
We construct a set of supersymmetric geometries that represent regular
microstates of the D1-D5-P 3-charge system, using the solution
generating technique of
\cite{Mathur:2003hj}.  These solutions are constructed as perturbations
around the maximally rotating D1-D5 solution at the linear order, and
depend on the coordinate of $S^1$ on which the D1- and D5-branes are
wrapped.  In the framework of six-dimensional supergravity developed by
Gutowski, Martelli and Reall
\cite{Gutowski:2003rg}, these solutions have a 4-dimensional base that
depend on the $S^1$ coordinate $v$.  The $v$-dependent base is expected
of the superstratum solutions which are parametrized by arbitrary
surfaces, and these solutions give a modest step toward their
explicit construction.

\newpage
\setcounter{page}{1} 





\section{Introduction}

The microphysics of black holes has been one of the most important
subjects in string theory which purports to be a consistent theory of
quantum gravity.  Since the pioneering work of Strominger and Vafa
\cite{Strominger:1996sh} on the supersymmetric D1-D5-P black hole, much
has been learned about the structures of black hole microstates.

The fuzzball conjecture \cite{Mathur:2005zp, Bena:2007kg, Mathur:2008nj,
Skenderis:2008qn, Balasubramanian:2008da, Chowdhury:2010ct} is about the
gravitational description of the black hole microstates.
The
conjecture claims that black hole microstates are made of
stringy/quantum gravity fuzz that extends over the horizon scale.
The example for which this conjecture is actually true is the
supersymmetric D1-D5 system (2-charge system).  For this system,
fuzzball microstates were explicitly constructed as smooth solutions in
classical supergravity, known as microstate geometries
\cite{Lunin:2001jy, Lunin:2002iz}, which were shown
\cite{Rychkov:2005ji} to correctly reproduce the asymptotic scaling of
the entropy expected from microscopic computation.  However, the
2-charge system is not really a black hole, the horizon area vanishing
classically.

The supersymmetric D1-D5-P system (3-charge black hole) has a finite
horizon and provides an ideal system in which to examine the fuzzball
conjecture.  The D1-D5-P system is obtained by compactifying type IIB
string theory on $S^1\times M_4$ with $M_4=T^4$ or K3, wrapping $N_1$
D1-branes on $S^1$ and $N_5$ D5-branes on $S^1\times M_4$, and putting
$N_p$ units of momentum along $S^1$.  Even if the fuzzball conjecture is
true, there is no a priori reason to expect that the black hole
microstates are describable in classical supergravity as smooth
solutions; they can be intrinsically stringy and have no supergravity
description at all.  Nonetheless, much effort has been made for
constructing microstate geometries for this system within supergravity
and, quite remarkably, many smooth solutions have been discovered.

In particular, a large family of smooth microstate geometries has been
explicitly constructed within supergravity in \cite{Bena:2005va,
Berglund:2005vb} (see also \cite{Lunin:2004uu, Giusto:2004id,
Giusto:2004ip} for earlier work).  This family can be characterized by
the fact that they are independent of the compact $S^1$ coordinate which
we call $v$.
Actually, however, there is growing evidence that this family is far
from the most generic microstates, even within supergravity.
As we mentioned above, the D1-D5-P system has momentum charge along $v$,
which can be naturally carried by traveling waves of the D1-D5
worldvolume depending on $v$ and, therefore, the corresponding solution
must be \emph{$v$-dependent}.  So, the family of $v$-independent
solutions in \cite{Bena:2005va, Berglund:2005vb} must not be the most
generic solutions.
Also, in \cite{Bena:2008nh, Bena:2008dw}, it was argued that placing
supertubes in the throat region of $v$-independent solutions can enhance
entropy.  This also suggests that $v$-dependence is important for
getting more generic solutions, because entropy of supertubes comes from
$v$-dependent fluctuations of the worldvolume which, upon backreaction,
turn $v$-independent background geometry into $v$-independent ones.
Furthermore, it has been shown that the $v$-independent solutions are
insufficient to account for the entropy of the D1-D5-P black hole
\cite{deBoer:2009un}.\footnote{Of course, it is fair to say that this
might instead be evidence that generic microstates are not describable
in supergravity.}
For these reasons, it is worthwhile to look for $v$-dependent microstate
solutions in supergravity in order to figure out whether the fuzzball
conjecture applies to the D1-D5-P system or not.  Considering
non-trivial dependence on the $S^1$ coordinate $v$ means that we must
consider six-dimensional solutions.

Some $v$-dependent solutions of supergravity have already been
constructed previously in the literature \cite{Mathur:2003hj,
Giusto:2006zi, Ford:2006yb, Mathur:2011gz, Mathur:2012tj, Lunin:2012gp,
Giusto:2012jx, Giusto:2013rxa} and were shown to represent smooth
microstates of the D1-D5-P system.
However, a systematic way to solve the relevant field equations in
general has not been found yet.  In this paper, we try to make a modest
progress in this direction, by studying $v$-dependent solutions in the
context of six-dimensional supergravity.
The supersymmetric solutions of this theory have been classified in
\cite{Gutowski:2003rg, Cariglia:2004kk} and, more recently, in Ref.\
\cite{Bena:2011dd}, the field equations that solutions should satisfy
have been recast into a form in which a linear structure is
manifest.\footnote{For recent applications of the linear structure for
constructing supergravity solutions, see \cite{Niehoff:2012wu,
Bobev:2012af, Vasilakis:2013tjs, Niehoff:2013kia}.  }$^,$\footnote{Ref.\
\cite{Giusto:2013rxa} discusses embedding of supersymmetric solutions in
a general class of 6D theory into 10D supergravity.} The solutions are
constructed based on a four-dimensional almost hyperk\"ahler base $\cB$
which can generally depend on $v$.  \emph{If the base $\cB$ is given}
and the source distribution of branes is given, all one has to do in
principle to obtain the backreacted solution is to solve the linear
system of differential equations.
However, the problem is that the base $\cB$ must satisfy certain
non-linear differential equations and we do not know how to solve them
in general. Namely, we lack a systematic method to construct the base $\cB$.

Most of the $v$-dependent solutions constructed thus far
\cite{Mathur:2003hj, Giusto:2006zi, Mathur:2011gz, Mathur:2012tj,
Lunin:2012gp, Giusto:2013rxa} have $v$-independent base $\cB$ (note
however the exceptions \cite{Ford:2006yb, Giusto:2012jx} which we
comment on below).  So, it is important to work out more explicit
examples of $v$-dependent base $\cB$ in detail, and that is what we will
do in this paper.

One may think that $v$-dependence of $\cB$ may not be crucial for
reproducing the correct scaling of the black hole entropy, just as for
the D1-D5 system where fluctuations in the $\bbR^4$ directions were
sufficient for the purpose of reproducing the entropy scaling and
fluctuations in the $T^4$ directions \cite{Kanitscheider:2007wq} were
not needed.
However, for the D1-D5-P system, there is an argument based on the
possibility of ``double bubbling'' that the $v$-dependence is essential
for getting the right entropy scaling.  This is a possibility that the
D1-D5-P system undergoes supertube transition multiple times
\cite{deBoer:2010ud, deBoer:2012ma} and its generic microstates are
represented by a brane configuration with $v$-dependent worldvolume,
dubbed the superstratum.  This double bubbling picture is supported by a
supersymmetry analysis \cite{Bena:2011uw}.  If this is true, we will
generically have a fluctuating distribution of KK monopoles
\cite{Bena:2011uw, Bena:2011dd} which is described by a $v$-dependent
base $\cB$, and we need to take them into account to reproduce the
entropy scaling.

In more detail, what we do in the current paper is to use the solution
generating technique \cite{Mathur:2003hj} to construct a solution with
$v$-dependent base.  In \cite{Mathur:2003hj}, they took the pure
AdS$_3\times S^3$ geometry which corresponds in boundary CFT to the NSNS
ground state. Around that background, they considered small fluctuation
of fields that corresponds to a chiral primary in CFT\@.  On the
fluctuation fields, they acted by a transformation which corresponds in
the bulk to a rotation in $S^3$ and which corresponds on the boundary to
an $R$-symmetry rotation.  This transformation changes the linear and
angular momenta carried by the fields.  Being just a rotation, this
transformation leaves smooth geometries smooth. After spectral flow to
the RR sector, this procedure gives a solution that carries
non-vanishing momentum charge.  Although they obtained a $v$-dependent
solution by this technique, their base was not $v$-dependent.
In this paper, we consider more general fluctuations around AdS$_3\times
S^3$ and apply their solution generating technique to obtain a
$v$-dependent base.\footnote{Note that all we do is an $S^3$ rotation
which is merely a coordinate transformation. So, in this sense, whether
the base is $v$-dependent or not is just a matter of the coordinate
system one uses.  However, what is important is that this coordinate
transformation does not vanish at the boundary of AdS$_3$.  This means
that this coordinate transformation generates genuinely new states in
the CFT, and that is what is important for microstate counting.}

It is appropriate here to mention the difference between our solution
and the solutions constructed in \cite{Ford:2006yb, Giusto:2012jx} which
also have $v$-dependent base.  Ref. \cite{Ford:2006yb} discussed
geometries obtained by the spectral flow of the Lunin-Mathur geometries
\cite{Lunin:2001jy, Lunin:2002iz} and correspond to CFT states on the
unitarity bound.  On the other hand, our solution is above the unitarity
bound and represent a different class of $v$-dependent solutions. Ref.\
\cite{Giusto:2012jx} constructed supergravity solutions by computing
perturbative open string amplitudes for certain brane bound states of
the D1-D5 system as the boundary states. This worldsheet-based method
has the advantage of being applicable to general boundary states but the
regularity of resulting solutions is difficult to study.  On the other
hand, in our approach, the regularity of the solution is easier to
analyze, although it is special to fluctuations around AdS$_3\times
S^3$.

Some comments on the relevance of smooth geometries for black hole
microstates are in order.  First, it is possible that a solution which looks
supersymmetric at the supergravity level may not be supersymmetric in
full string theory \cite{Dabholkar:2009dq, Chowdhury:2013pqa} (see also
\cite{Chang:2013fba}).  So, a given supergravity solution might not
actually represent a microstate of the black hole in question.
Second, the analysis of quiver quantum mechanics \cite{Bena:2012hf}
representing multi-center black holes in 4D suggests that the black hole
microstates may correspond to ``pure Higgs'' states with vanishing
angular momentum, which is rather unnatural from the viewpoint of
microstate geometries.
Note that these two facts are \emph{not} necessarily pointing toward the
irrelevance of microstate geometries for the fuzzball conjecture; it may
instead be completely opposite.
Namely, it seems natural to interpret them as saying that microstate
geometries are generally lifted by an amount invisible in supergravity
except for ones with vanishing angular momentum.  This would nicely
explain the fact that the angular momentum of supergravity microstates
is not restricted to zero whereas quiver quantum mechanics suggests that
the truly supersymmetric states have vanishing angular momentum.
Further investigations are needed to clarify the relevance of microstate
geometries, including the ones constructed in the current paper, as the
true microstates of the supersymmetric D1-D5-P black hole.  In
particular, $v$-dependent solutions are expected to play an important
role.

The organization of the rest of the paper is as follows.  In section
\ref{sec:review_6d}, we review the supersymmetric solutions in the
six-dimensional supergravity theory of our interest, and how they can be
embedded in 10D supergravity.  After reviewing the solution generating
technique of \cite{Mathur:2003hj} in section \ref{sec:MSS_soln_gen}, we
present the construction of the solution in section
\ref{sec:construction}.  We will only describe the outline and the
result, referring to the appendix for details.  In section
\ref{sec:future}, we discuss possible future directions.

\section{Review of supersymmetric solutions in 6D}
\label{sec:review_6d}

Here we review the supersymmetric solutions in 6D supergravity as
presented in \cite{Bena:2011dd}.  We will be brief here; for more
details the reader is referred to \cite{Gutowski:2003rg,
Cariglia:2004kk, Bena:2011dd}.

The classification of supersymmetric solutions in 6D $\cN=1$
supergravity was first done by Gutowski, Martelli, and Reall (GMR)
\cite{Gutowski:2003rg} for minimal supergravity and later generalized in
\cite{Cariglia:2004kk} to include vector multiplets.  The supergravity
theory we consider here is $\cN=1$ theory with an anti-self-dual tensor
multiplet  \cite{Bena:2011dd}, and its bosonic field content consists of the metric
$g_{\mu\nu}$, an unconstrained 2-form $B_2$ with field strength
$G=dB_2$, and a dilaton $\phi$.
The most general supersymmetric solutions for this theory have a null
Killing direction $u$, of which all fields are independent.  However,
the fields can in general depend on the remaining five coordinates.
Because null Killing vector introduces a $2+4$ split in the geometry, it
is natural to introduce a second retarded time coordinate $v$ and a
four-dimensional, and generically $v$-dependent, spatial base $\cB$
with coordinates $x^m$, $m=1,\dots,4$.

The six-dimensional metric is given by
\begin{align}
 ds_6^2&=2H^{-1}(dv+\beta)\,\bigl[du+\omega+{1\over 2}\cF(dv+\beta)\bigr]
 -Hds_4^2,\label{GMRmetric}
\end{align}
where $H,\cF$ are functions and $\beta,\omega$ are 1-forms in $\cB$.
$H,\cF,\beta,\omega$ in general depend on $v,x^m$.  The base $\cB$ has
the metric
\begin{align}
 ds_4^2=h_{mn}dx^m dx^n
 \label{Bmetric}
\end{align}
and equipped with almost hyperk\"ahler structure 2-forms $J^{(A)}$, $A=1,2,3$,
which are anti-self-dual,
\begin{align}
 *_4 J^{(A)}=-J^{(A)},\label{J_ASDty}
\end{align}
and satisfy the quaternionic relation
\begin{align}
 J^{(A)}{}^m{}_n J^{(B)}{}^n{}_l
 =-\delta^{AB}\delta^m_l+\epsilon^{ABC}J^{(C)}{}^m{}_l,\qquad
 J^{(A)}{}^m{}_n\equiv g^{ml}J^{(A)}_{ln}.\label{J_quat}
\end{align}
Here, $*_4$ is the Hodge star with respect to the four-dimensional
metric \eqref{Bmetric}.  For our convention of differential forms and
Hodge star, see Appendix \ref{app:conv}\@.
The 2-forms $J^{(A)}$ are not closed but its non-closure is
related to $\beta$ as
\begin{align}
 \tilde d J^{(A)}&=\partial_v (\beta\wedge J^{(A)}),\label{J_diff_cond}
\end{align}
with $\tilde d$ being the exterior derivative restricted to the base,
$\tilde d=dx^m \partial_m$.  The 1-form $\beta$ must satisfy the
condition
\begin{align}
 D\beta=*_4 D\beta\label{beta}
\end{align}
where
\begin{align}
 D\equiv \tilde{d}-\beta\wedge \partial_v~.
\end{align}
We also introduce the 2-form
\begin{align}
 \psih\equiv {1\over 16}\epsilon^{ABC}J^{(A)ij}\dot J^{(B)}_{ij}J^{(C)},
\label{psidef}
\end{align}
which measures the rotation of $J^{(A)}$ as $v$ varies.  Here, we
defined $ \dot{~}\equiv \partial_v$.

Given the base $\cB$ and the 1-form $\beta$ satisfying the above
equations, we can determine $H,\omega$ in the metric, the dilaton
$\phi$, and the flux $G=dB_2$ by solving a linear system as follows.
We introduce functions $Z_1,Z_2$  by
\begin{align}
 Z_1=He^{\sqrt{2}\phi},\quad
 Z_2=He^{-\sqrt{2}\phi},
 \label{Zdef}
\end{align}
and $2$-forms $\Theta_1,\Theta_2$.
Then they satisfy the following linear equations:
\begin{align}
 \begin{split}
 D*_4(DZ_1+\dot\beta Z_1)&=-2\Theta_2\wedge D\beta,\qquad
 \tilde d\Theta_2=\partial_v\left[{1\over 2}*_4(DZ_1+\dot\beta Z_1)+\beta\wedge \Theta_2\right]
 ,
 \\
 D*_4(DZ_2+\dot\beta Z_2)&=-2\Theta_1\wedge D\beta,\qquad
 \tilde d\Theta_1=\partial_v\left[{1\over 2}*_4(DZ_2+\dot\beta Z_2)+\beta\wedge \Theta_1\right]
 \end{split}
\label{EOM-Bianchi}
\end{align}
The $\Theta_{1,2}$ are not quite self-dual but the failure is related to $\psih$ as
\begin{align}
 *_4\Theta_1&=\Theta_1-2Z_2\psih,
 \qquad
 *_4\Theta_2=\Theta_2-2Z_1\psih.
  \label{EOM-Bianchi_Theta_duality}
\end{align}
Once $Z,\Theta$ are known, the field strength $G=dB_2$ is given by
\begin{align}
 G&=d\!\left[-{1\over 2}Z_1^{-1}(du+\omega)\wedge (dv+\beta)\right]+\widehat G_1,\label{gxql26Aug11}\\
 e^{2\sqrt{2}\phi}*_6 G
 &=
 d\!\left[-{1\over 2}Z_2^{-1}(du+\omega)\wedge (dv+\beta)\right]+\widehat G_2,
 \label{heda14Mar12}
\end{align}
where
\begin{align}
 \widehat G_1&\equiv {1\over 2}*_4 (D+\dot \beta)Z_2+(dv+\beta)\wedge \Theta_1,\label{lojy24Jun13}\\
 \widehat G_2&\equiv {1\over 2}*_4 (D+\dot \beta)Z_1+(dv+\beta)\wedge \Theta_2.
\end{align}
The 1-form $\omega$ is found by solving the equation
\begin{align}
 (1+*_4)D\omega&=
 2(Z_1\Theta_1+Z_2\Theta_2)-\cF D\beta-4Z_1Z_2\psih.
 \label{omegaEOM}
\end{align}
Finally, $\cF$ is determined by 
\begin{align}
 *_4D*_4L
 &={1\over 2}Hh^{ij}\partial_v^2(Hh_{ij})
 +{1\over 4}\partial_v(Hh^{ij})\partial_v(Hh_{ij})
 \notag\\
 &\qquad\qquad
 -2\dot\beta_i L^i+2H^2\dot\phi^2
 -2*_4[\Theta_1\wedge \Theta_2-\psih\wedge D\omega],
 \label{LEOM}
\end{align}
where
\begin{align}
 L&\equiv \dot \omega+{1\over 2}\cF\dot \beta-{1\over 2}D\cF.
 \label{Ldef}
\end{align}

We embed the above 6D theory into 10D type IIB supergravity as
follows \cite{Kanitscheider:2006zf, Kanitscheider:2007wq} (note that 
embedding is not unique).  We identify the 2-form $B_2$ with the RR
2-form potential $C_2$ and the 6D dilaton $\phi$ with the 10D dilaton
$\Phi$ as
\begin{align}
 B_2={1\over 2}C_2,\qquad
 \phi={1\over \sqrt{2}}\Phi.\label{B2_and_C2}
\end{align}
Then the relation of $G=dB_2$ to the RR 3-form flux $F_3=dC_2$ and the dual
$F_7=*_{10}F_3$ is
\begin{align}
 G&=\thalf F_3=\thalf dC_2,
 \qquad
 e^{2\sqrt{2}\phi}\,{*_6 G}
 = \thalf          F_7 |_6.\label{iorb7Oct11}
\end{align}
Here, $[\dots]|_6$ means to strip off the $M_4$ part of the differential
form.
Because $F_3\propto G$ couples electrically to D1 and magnetically to
D5, the first term $d[...]$ of $G$ in \eqref{gxql26Aug11} corresponds to
D1$(u,v)$ and the function $Z_1$ is the potential for it.  The first
term of $\Gh_1$ in \eqref{lojy24Jun13} corresponds to D5$(u,v,M_4)$ and
the second term in $\Gh_1$ to D5$(u,\psi,M_4)$ where $\psi$ is some
curve in $\cB$.  Inside $\cB$, D5$(u,\psi,M_4)$ is a 1-brane along
$\psi$ and we can measure its charge by integrating $\Theta_1$ over a
2-surface going around it.  From $e^{2\sqrt{2}\phi}*_6G\propto F_7$, we
can similarly read off charges, setting D1$\leftrightarrow$D5.
Also, $\beta_m,\omega_m$ correspond to linear combinations of momentum
charge along $x^m$ and KK monopole charge along $x^m\times M_4$ with
special circle $v$.

\section{Solution generating technique}
\label{sec:MSS_soln_gen}

In this section, we review the solution generating technique by Mathur,
Saxena, and Srivastava (MSS) \cite{Mathur:2003hj}, which allows one to
construct a solution carrying momentum charge starting with a seed
solution carrying no momentum charge.

The Lunin-Mathur (LM) geometry \cite{Lunin:2001jy, Lunin:2002iz} is a
family of smooth geometries in 6D describing microstates of the D1-D5
system.  They are parametrized by continuous functions $F_m(w)$ called
the profile function which parametrizes the closed curve in $\bbR^4$
along which the D1-D5 worldvolume is extending.\footnote{We do not
discuss the generalization for the profile function to describe
fluctuations in the $T^4$ directions \cite{Kanitscheider:2007wq}.} They
represent the ground states in the RR sector of the D1-D5 CFT and the
dictionary between the geometries and CFT states is well established
\cite{Kanitscheider:2006zf}.  Expressed in the GMR form of section \ref{sec:review_6d}, the
LM geometry is given by the following $v$-independent functions and
forms \cite{Ford:2006yb}:
\begin{align}
\begin{split}
 Z_1&={Q_5\over L}\int_0^L {|\dot{\vec{F}}(w)|^2dw\over|\vec{x}-\vec{F}(w)|^2},\qquad
 Z_2={Q_5\over L}\int_0^L {dw\over|\vec{x}-\vec{F}(w)|^2},\\
 ds_4^2& = \delta_{mn} dx^m dx^n,\qquad
 \cF=0,\qquad \beta=-{A+B\over\sqrt{2}},\qquad\omega=-{A-B\over\sqrt{2}},\\
 A_m&=-{Q_5\over L}\int_0^L { \dot F_m(w)dw\over|\vec{x}-\vec{F}(w)|^2},\qquad
 dB = *_4 dA,\qquad
 \Theta_1=\Theta_2=\psih=0.
\end{split}\label{LM_in_GMR_form}
\end{align}
where $L$ is a constant defined in \eqref{L_def}, $Q_5$ is the D5 charge
proportional to $N_5$ (see \eqref{Q_and_N}) and the D1 charge $Q_1$ is
given in \eqref{Q1_rel}.  The profile function satisfies the periodicity
condition $F_m(w+L)=F_m(w)$.  The RR 2-form $C_2$, which is related to the
2-form $B_2$ by \eqref{B2_and_C2}, is given by
\begin{align}
 C_2
 &=-Z_1^{-1}
  (du+\omega)\wedge(dv+\beta)
 +\cC_2,\qquad d\cC_2=*_4dZ_2,
 \label{C2_LM}
\end{align}
which is nothing but \eqref{gxql26Aug11}, \eqref{lojy24Jun13}.  Note
that we dropped ``1'' in the harmonic functions $Z_{1,2}$ so that the
above solution describes asymptotically AdS space.  Extending our
computation to asymptotically flat space would be interesting but we
will not attempt to do it in this paper.
See Appendix \ref{app:LM_geom} for more about the LM geometry.

In \cite{Mathur:2003hj}, MSS
constructed a $v$-dependent 3-charge configurations by considering small
fluctuations around maximally rotating LM geometry
\cite{Balasubramanian:2000rt, Maldacena:2000dr}.  This geometry is given by a circular profile function,
\begin{align}
 F_1+iF_2&=ae^{i\omega w},\qquad
 F_3=F_4=0,\qquad \omega={2\pi\over L}
 ,\qquad
 a={\sqrt{Q_1Q_5}\over R},
 \label{MRLM_profile}
\end{align}
with $R$ being the radius of $S^1$, and represents a particular RR
ground state of the D1-D5 CFT with maximal possible $R$-charge.  In this
case, the GMR data \eqref{LM_in_GMR_form} and \eqref{C2_LM} are computed
to be
\begin{align}
 \begin{split}
 Z_{1}&={Q_{1}\over h},\qquad
 Z_{2}={Q_{5}\over h},\qquad
 h\equiv r^2+a^2\cos^2\theta,
 \\
 A&=-{\sqrt{Q_1Q_5}\,a\,\sin^2\theta\over h}d\phi,\qquad
 B={\sqrt{Q_1Q_5}\,a\,\cos^2\theta\over h}d\psi,\\
 \cC_2&=-{Q_5(r^2+a^2)\cos^2\theta\over h} d\phi\wedge d\psi.
 \end{split}
 \label{GMRdata_circular_profile}
\end{align}
Here, we introduced the coordinates $r,\theta,\phi,\psi$  by
\cite{Lunin:2002iz}
\begin{align}
\begin{split}
 &x^1+ix^2=se^{i\phi},\qquad
 x^3+ix^4=we^{i\psi},\\
 &s=\sqrt{r^2+a^2\,}\,\sin\theta,\qquad
 w=r\cos\theta,\\
 s,w,r&\in[0,\infty),\qquad 
 \phi,\psi\in[0,2\pi),\qquad
 \theta\in[0,{\pi\over 2}],
\end{split}
\end{align}
in terms of which the metric for the flat 4D base becomes
\begin{align}
 ds_4^2
 &=
 h\left({dr^2\over r^2+a^2}+d\theta^2\right)
 +(r^2+a^2)\sin^2\theta \,d\phi^2
 +r^2\cos^2\theta\, d\psi^2.\label{ds42_flat}
\end{align}

By the spectral flow
transformation of the CFT, this state can be mapped into the ground
state in the NS-NS sector. In the bulk, the spectral flow corresponds to
a simple coordinate transformation
\begin{align}
 \phit=\phi-{t\over R},\qquad
 \psit=\psi+{y\over R},
 \label{spectralflow}
\end{align}
where
\begin{align}
 t={u+v\over \sqrt{2}},\qquad
 y={u-v\over \sqrt{2}}.\label{fjqx21Jun12}
\end{align}
One can show that this brings the 6D metric \eqref{GMRmetric} into
AdS$_3\times S^3$:
\begin{subequations}
 \begin{align}
  ds^2_{6}&=-ds^2_{\rm AdS_3}-\sqrt{Q_1Q_5}\,ds^2_{S^3},\\
 ds^2_{\rm AdS_3}&={1\over\sqrt{Q_1Q_5}}
 \left[-(r^2+a^2)dt^2+r^2dy^2+{Q_1Q_5\over r^2+a^2}dr^2\right],
   \label{AdS3_metric}\\
 ds^2_{S^3}&=d\theta^2+\sin^2\theta\, d\phit^2+\cos^2\theta\, d\psit^2,
   \label{S3_metric}\\
 C_2&={r^2+a^2\over Q_1}dt\wedge dy
 +\sqrt{Q_5\over Q_1}\,{a}\,d\phit\wedge dy
 -Q_5\cos^2\theta\, d\phit\wedge d\psit.
 \end{align}
 \label{AdS3xS3}
\end{subequations}
Around this AdS$_3\times S^3$ background, MSS considered a fluctuation of
the fields that corresponds to a chiral primary with
\begin{align}
 (h^{\rm NS},j^{\rm NS})=(k,k),\qquad
 (\bar{h}^{\rm NS},\bar{\jmath}^{\rm NS})=(k,k),
 \label{ipn25Jun13}
\end{align}
where $h,\bar{h}$ are the eigenvalues of the Virasoro generators
$L_0,\bar{L}_0$ while $j,\bar{\jmath}$ are the eigenvalues of the
$SU(2)\times \widetilde{SU(2)}$ $R$-symmetry generators $J^3_0,\bar{J}^3_0$.  The
subscript NS denotes the NS sector.  The corresponding bulk fields can
be worked out using the field equations of 6D supergravity.
If one did the inverse spectral flow transformation to this state, the
one would obtain an RR ground state which has less than  maximal
$R$-charge and no momentum charge.  In order to generate a new solution,
they instead acted by $(J_0^-)^{\rm NS}$ on the state \eqref{ipn25Jun13}
to get an NSNS state with
\begin{align}
 (h^{\rm NS},j^{\rm NS})=(k,k-1),\qquad
 (\bar{h}^{\rm NS},\bar{\jmath}^{\rm NS})=(k,k),
 \label{wuz25Jun13}
\end{align}
and then did the inverse spectral flow.  In the bulk, $(J_0^-)^{\rm NS}$
corresponds to one of the generators of the
$SO(4)=SU(2)\times\widetilde{SU(2)}$ rotation group of $S^3$ and is represented by a
simple differential operator. So, it is easy to work out the fields
corresponding to \eqref{wuz25Jun13}.  After inverse spectral flow
transformation\footnote{Note that this is for the weight and $R$-charge
of the perturbation, not including that of the background.}
\begin{align}
 h^{\rm R}=h^{\rm NS}-j^{\rm NS},\qquad j^{\rm R}=j^{\rm NS},
\end{align}
we end up with an RR state with
\begin{align}
 (h^{\rm R},j^{\rm R})=(1,k-1),\qquad
 (\bar{h}^{\rm R},\bar{\jmath}^{\rm R})=(0,k),
\end{align}
which has non-vanishing momentum charge
\begin{align}
 N_p=
 h^{\rm R}-\bar{h}^{\rm R}=1.
\end{align}
Being a simple $SU(2)$ rotation of the original solution, this solution
is guaranteed to be smooth and represents a microstate of the D1-D5-P
system.

MSS studied particular chiral primaries which are represented in 6D
supergravity \cite{Deger:1998nm}\footnote{There are different ways to
embed the 6D fields into 10D fields, and they correspond to different
chiral primaries.  For particular ways to embed solutions in 6D
supergravity into 10D supergravity, see e.g.\
\cite{Kanitscheider:2006zf, Kanitscheider:2007wq, Giusto:2013rxa}.} by
fluctuations only of 6D dilaton and gauge fields but does not change the
background metric from AdS$_3\times S^3$.\footnote{This is true only at
the first order in the fluctuation.  At higher order, the fields
backreact on the metric and the background will change.}  The latter
fact greatly simplified their analysis but at the same time implies
that, when recast in the GMR form, the solution has a $v$-independent
base.

\section{Construction of the $v$-dependent solution}
\label{sec:construction}

In this section, we use the solution generating technique reviewed above
to construct a 3-charge solution with $v$-dependent base $\cB$. Here we
will outline the main computations, followed by a summary of the
results, relegating some details to Appendix \ref{app:detail}.

\subsection{The seed solution and  spectral flow}
\label{ss:seed_and_spectral_flow}

We would like to use the solution generating technique of MSS reviewed
above in order to obtain a solution with a $v$-dependent base.  For
that we need fluctuations
more general than was considered by MSS\@.
Specifically, as the ``seed'', we take the following fluctuation of the
LM profile,
\begin{align}
 \delta F_1+i\,\delta F_2=be^{i(k+1)\omega w+i\alpha},\qquad
 \delta F_3=\delta F_4=0,\label{bjga5Jul13}
\end{align}
around the circular profile \eqref{MRLM_profile}.  Here, $b$ is a small
number and we will work only at the linear order in expansions in $b$.
$\alpha$ is an arbitrary constant phase while
\begin{align}
 k\in\bbZ,\qquad k\le -2\quad \text{or}\quad  1\le k.
\label{krange}
\end{align}
$k=-1$ is excluded because it would correspond to translating the entire
profile, while $k=0$ is excluded because it would correspond to changing
the background radius $a$ and change the D1 charge $Q_1$.  The change in
the GMR data, such as $\delta Z_1$, can be computed readily by plugging
$F+\delta F$ into \eqref{LM_in_GMR_form} and expanding it in the small
parameter $b$ (see \eqref{dZ_and_dA}).
Actually, it is more convenient to take a suitable linear combination of
fluctuations with different phase $\alpha$, which we are permitted to do in the
linear approximation.  Specifically, taking the linear combination
$(\alpha=0)+i(\alpha=-\pi/2)$, we find that the change in the GMR data is
\begin{align}
\begin{split}
  \delta Z_1&=2Q_5 ab\,\omega^2
 \left[a(sI_2(k+1)-aI_2(k))+ (k+1)I_1(k)\right]e^{ik\phi},
 \\
 \delta Z_2&=2Q_5 b\,(sI_2(k+1)-aI_2(k))e^{ik\phi},\\
 \delta A&=Q_5b\omega\, (- iX_-\,ds-sX_+\,d\phi)e^{ik\phi},\qquad
 \delta B=2Q_5 ab\omega w^2 e^{ik\phi}I_2(k)d\psi,
\end{split}
  \label{dGMRdata}
\end{align}
where $I_n(k),X_\pm$ are defined in \eqref{I_n(k)_def}, \eqref{Xpm_def}.
$\Theta_{1,2},\cF,\psih$ still vanish, because we are dealing with the
LM geometry anyway.\footnote{We can identify the fluctuation studied in
MSS \cite{Mathur:2003hj}, which does not change the 6D metric, with a
linear combination of the fluctuation \eqref{dGMRdata}.  Specifically,
if we denote the fields in \eqref{dGMRdata} depending on $k$
collectively by $F(k)$, then the fluctuation in \cite{Mathur:2003hj}
corresponds to $\half(F(k)-F(-k)^*)$.  In terms of the profile function
$F_m(w)$, this is a ``longitudinal'' fluctuation that does not change
the shape but only the parametrization.  More precisely, one can show
that it corresponds to $(F_1+iF_2)+(\delta F_1 +i \delta F_2)=a\exp[i\omega
(w+(b/a\omega)\sin(k\omega w))]$.}

The GMR data \eqref{dGMRdata} represent a small fluctuation around the
maximally rotating LM geometry.  This solution still belongs to the LM
geometries \eqref{LM_in_GMR_form} and therefore corresponds to a certain
RR ground state of the D1-D5 CFT\@.  To use the solution generating
technique of MSS, let us do a spectral flow transformation to the NS
sector, so that we have fluctuating fields around AdS$_3\times S^3$.  To
the zeroth order, the spectral flow transformation is implemented by the
coordinate transformation \eqref{spectralflow} but, in the presence of
the fluctuation on top, we have the freedom to do a further coordinate
transformation at the same order in $b$.  Let us use this
freedom to bring the fluctuation of the metric into the canonical form
of Deger et al.\ \cite{Deger:1998nm}.  Concretely, we apply the
following coordinate transformation\footnote{Part of this coordinate
transformation has been written down in \cite{Donos:2005vs}.  This is a
generalization so that the full 6D metric is in the form given in
\cite{Deger:1998nm}, not just the $S^3$ part.}
\begin{align}
\begin{split}
  \xi^\mu &=(\xi^t,\xi^y,\xi^r,\xi^\theta,\xi^\phit,\xi^\psit)\\
 &={b a^{|k|} e^{ik({t/ R}+\phit)}\sin^{|k|}\theta\over (r^2+a^2)^{|k|/2}}
 \left(\mp i{\sqrt{Q_1Q_5}\over r^2+a^2},0,{ar\sin^2\theta\over h},
 {a\sin\theta\cos\theta\over h},0,0\right),\\
 g_{\mu\nu}&\to g_{\mu\nu}+\nabla_\mu\xi_\nu+\nabla_\nu\xi_\mu,
\end{split}\label{bjoy5Jul13}
\end{align}
where the $\mp$ signs correspond to $k\gtrless 0$, respectively.  Then
the change in the 6D metric, relative to the AdS$_3\times S^3$ metric
\eqref{AdS3xS3}, takes a rather simple form as follows:
\begin{multline}
 \delta(ds_6^2)=(|k|+1) \,a^{|k|-1}b\, \Bh\, \Yh
 \biggl[
 {r^2-a^2\over \sqrt{Q_1Q_5}}dt^2-{r^2dy^2\over \sqrt{Q_1Q_5}}
 +{\sqrt{Q_1Q_5}(r^2-a^2)dr^2\over (r^2+a^2)^2}
 \\
 \mp{4iar\over r^2+a^2}dtdr
 +\sqrt{Q_1Q_5}(d\theta^2+\sin^2\theta d\phit^2+\cos^2\theta d\psit^2)
 \biggr],\label{delta_ds62}
\end{multline}
where
\begin{align}
 \Bh\equiv { e^{ikt/R}\over (r^2+a^2)^{|k|/2}}
 ,\qquad
 \Yh\equiv e^{ik\phit}\sin^{|k|}\theta.
 \label{def_B,Y}
\end{align}
We can also find the change in dilaton to be
\begin{align}
  \delta \Phi= \sqrt{2}\,\delta \phi
 &=(k+1)a^{|k|-1}b\Bh\Yh.\label{dilatonnonlong}
\end{align}
Also, the change in the RR 2-form relative to \eqref{AdS3xS3} can be written
in the canonical form of \cite{Deger:1998nm} as
\begin{align}
\delta C_2=
\begin{cases}
 \displaystyle -{2 (k+1) a^{k-1} b \over Q_1 \omega}
 \Bh\Yh
\left[r^2 \omega\,  dt\wedge dy+i Q_1 \frac{r\, dy\wedge dr}{r^2+a^2}\right]
 &  \qquad (k>0),
 \\[3ex]
 \displaystyle  -{2 a^{l-1} b \over Q_1 \omega}\Bh\Yh\biggl[r^2 \omega\,  dt\wedge dy
 -i Q_1 \frac{r\, dy\wedge dr}{r^2+a^2}
 \\
 \displaystyle \qquad\qquad
 -ilQ_1Q_5\omega \cot\theta
 (d\theta -i\sin\theta\cos\theta d\phit)\wedge d\psit\biggr]
 & \qquad (k=-l<0).
\end{cases}
\label{delta_C_2}
\end{align}
See Appendix \ref{app:seed_spectral_flow} for details.

\subsection{$\boldsymbol{SU(2)}$ rotation}

Now we would like to do a transformation to the fluctuation
\eqref{delta_ds62}, \eqref{dilatonnonlong}, \eqref{delta_C_2} to
generate a new solution.  The $S^3$ is parametrized by
$\theta,\psit,\phit$, and its isometry group $SO(4)=SU(2)\times \widetilde{SU(2)}$ is
generated by\footnote{
The $SO(4)$ generators
$J^{mn}=-i(x_m\partial_n-x_n\partial_m)$, $m,n=1,2,3,4$ can be split into
$SU(2)\times \widetilde{SU(2)}$ generators as
$J^a=J_{+}^{a4}$, $\bar{J}^a=J_{-}^{a4}$, $a=1,2,3$, where
$J_\pm^{mn}={1\over 2}(\tilde
J^{mn}\pm J^{mn})$, $\tilde J^{mn}={1\over 2}\epsilon_{mnpq}J^{pq}$.
}
\begin{align}
\label{eq:JL,JR}
\begin{split}
  J^\pm &={i\over 2}e^{\pm i(\phit+\psit)}
 (\mp i\partial_\theta+\cot\theta\, \partial_\phit-\tan\theta\, \partial_\psit)
 ,\qquad
 J^3=-{i\over 2}(\partial_\phit+\partial_\psit),\\
 \bar{J}^\pm &={i\over 2}e^{\pm i(\phit-\psit)}
 (\mp i\partial_\theta+\cot\theta\, \partial_\phit+\tan\theta\, \partial_\psit)
 ,\qquad
 \bar{J}^3=-{i\over 2}(\partial_\phit-\partial_\psit).
\end{split}
\end{align}
For $k>0$, all the fluctuation fields \eqref{delta_ds62},
\eqref{dilatonnonlong}, \eqref{delta_C_2} are proportional to the scalar
spherical harmonic with the highest weight $(k,k;k,k)$ of $SU(2)\times
\widetilde{SU(2)}$,
\begin{align}
  \Yh={e^{ik\phit}\sin^{k}\theta},\qquad k>0,
\end{align}
which is killed by $J^+,\bar{J}^+$. 
This means that the fluctuation fields have
\begin{align}
 (h^{\rm NS},j^{\rm NS})=(k,k),\qquad
 (\bar{h}^{\rm NS},\bar{\jmath}^{\rm NS})=(k,k).
\end{align}
Since the background preserves the $SU(2)\times \widetilde{SU(2)}$
symmetry, the above solution remains a solution even if we replace $\Yh$
with the $(k,k-m;k,k)$ state,
\begin{align}
   (J^-)^m \Yh\propto {e^{i(k-m)\phit-im\psit}\sin^{k-m}\theta\cos^m \theta}
 \equiv \Yt,
\end{align}
which has
\begin{align}
 (h^{\rm NS},j^{\rm NS})=(k,k-m),\qquad
 (\bar{h}^{\rm NS},\bar{\jmath}^{\rm NS})=(k,k).
\end{align}
After this replacement $\Yh\to\Yt$, we go back to the RR sector by the
spectral flow transformation \eqref{spectralflow}.  (Note that we do not
do a coordinate transformation similar to \eqref{bjoy5Jul13} before
spectral flowing back.) The resulting configuration has
\begin{align}
 (h^{\rm R},j^{\rm R})=(m,k-m),\qquad
 (\bar{h}^{\rm R},\bar{\jmath}^{\rm R})=(0,k)
\end{align}
and therefore the momentum charge
\begin{align}
 N_p=h^{\rm R}-\bar{h}^{\rm R}=m.
\end{align}
The resulting fields can be rewritten in the GMR form, as summarized in
the next subsection.  

For $k=-l<0$, on the other hand, the fields are proportional to
\begin{align}
 \Yh=e^{-il\phit}\sin^l\theta,\qquad l>0,
\end{align}
which is the lowest state $(l,-l;l,-l)$.  The corresponding CFT charges are
\begin{align}
 (h^{\rm NS},j^{\rm NS})=(l,-l),\qquad
 (\bar{h}^{\rm NS},\bar{\jmath}^{\rm NS})=(l,-l).
\end{align}
Acting on the state by $(J^+)^n$, $n>0$, we obtain the $(l,-(l-n);l,-l)$
state
\begin{align}
 \Yt\propto (J^+)^n \Yh \propto
 {e^{i(-l+n)\phit+in\psit}\sin^{l-n}\theta\cos^n \theta}.
\end{align}
After inverse spectral flow, we end up with an RR state with
\begin{gather}
 (h^{\rm R},j^{\rm R})=(2l-n,-l+n),\qquad
 (\bar{h}^{\rm R},\bar{\jmath}^{\rm R})=(2l,-l),\\
 N_p=h^{\rm R}-\bar{h}^{\rm R}=-n.
\end{gather}

The expression for $\Yt$ that works for both $k>0,k<0$ is
\begin{align}
 \Yt={e^{i(k-m)\phit -im\psit}\sin^{|k|-|m|}\theta\cos^{|m|} \theta},
\end{align}
where for $k<0$ we take $m=-n<0$. The value of $m$ is restricted to
$0\le |m|\le |k|$.

\subsection{The $\boldsymbol{v}$-dependent solution}

As the result of the procedure outlined above, we obtain the following
GMR fields representing a microstate of the D1-D5-P system:
\begin{subequations}
\begin{align}
\delta H&={c\sqrt{Q_1Q_5}(r^2-a^2\cos^2\theta)\over h^2}F,\qquad
\sqrt{2}\,\delta \phi=\delta \Phi=(k+1) a^{|k|-1}b F
 \\
 \delta Z_1&=  {a^{|k|-1} b Q_1 \over h^2}\left[r^2(k+|k|+2)+a^2(k-|k|)\cos^2\theta\right] F,\\
 \delta Z_2&= -{a^{|k|-1} b Q_5 \over h^2}\left[r^2(k-|k|)+a^2(k+|k|+2)\cos^2\theta\right]F,
\\
 \delta\beta&={ac\sqrt{2Q_1Q_5}\over h}F
  \left[\pm{ir\,dr\over r^2+a^2}+{r^2\over h}(\sin^2\theta d\phi -\cos^2\theta d\psi)\right],\\
 \delta\omega&={ac\sqrt{2Q_1Q_5}\over h}F
 \left[\pm{ir\,dr\over r^2+a^2}+{r^2\over h}(\sin^2\theta d\phi +\cos^2\theta d\psi)\right],\\
 \cF&=0,\\
 \delta( ds_4^2)&= {2 a^2 c F}
 \left[
 \sin^2\theta\,\left(d\phi\pm{ir\,dr \over r^2+a^2} \right)^2+\cos^2\theta \,d\theta^2 \right],
\end{align}
\label{dGMR_data_final}
\end{subequations}
where
\begin{align}
 c\equiv (|k|+1)a^{|k|-1} b,\qquad 
 F\equiv {e^{i{\sqrt{2}}mv/R+i(k-m)\phi-im\psi }\sin^{|k|-|m|}\theta\cos^{|m|} \theta\over
 (r^2+a^2)^{|k|/2}
 }.
 \label{def_F}
\end{align}
Here $k\in\bbZ$ ($k\neq -1,0$) and $|m|\le |k|$. The sign of $m$ is also
correlated to that of $k$, namely, $\sign(m)=\sign(k)$.  The $\pm$ signs
above correspond to $k\gtrless 0$.  We can see that the base metric is
$v$-dependent as we wanted.  This solution carries non-vanishing momentum
\begin{align}
 N_p=m.
\end{align}

Note that, in our approximation at first order in perturbation, we have
$\cF=0$ and we cannot read off the momentum charge from the asymptotic
behavior of $g_{uv}$.  This is because the metric starts to feel
momentum only at the quadratic order, because the energy-momentum tensor
$T_{\mu\nu}$ is quadratic in fields.

The $\Theta$ fields can be read off from \eqref{gxql26Aug11} and
\eqref{heda14Mar12} as
\begin{align}
 \delta \Theta_1&=
 \begin{cases}
 (k+1)m a^{k+2}b\sqrt{2Q_5\over Q_1}
 {F\over h^2}\cos^2\theta
 \left( {rh\over r^2+a^2}dr-ir^2\sin^2\theta\, d\phi \right)\wedge d\psi
  &(k>0)
 \\[3ex]
 |m| a^{|k|}b\sqrt{2Q_5\over Q_1}F
 \biggl[ |k|\tan\theta \left(-{ir\,dr\over r^2+a^2}+d\phi\right)\wedge d\theta
  \\
 \qquad\qquad\qquad\qquad
 +{(-|k|r^2+a^2\cos^2\theta)r\over h}\left({dr\over r^2+a^2}+{ir\over h}\sin^2\theta d\phi\right)\wedge d\psi\biggr],&(k<0).
 \end{cases}
 \label{dTheta1_final}
 \\
 \delta \Theta_2&=
 \begin{cases}
 (k+1)m a^{k}b\sqrt{2Q_5\over Q_1}F
 \biggl[ \tan\theta \left({ir\,dr\over r^2+a^2}+d\phi\right)\wedge d\theta
  \\
 \qquad\qquad\qquad\qquad
 +{r^3\over h}\left(-{dr\over r^2+a^2}+{ir\over h}\sin^2\theta d\phi\right)\wedge d\psi\biggr]&(k>0).
 \\[3ex]
 |m| a^{|k|}b\sqrt{2Q_5\over Q_1}F
 \biggl[ \tan\theta \left(-{ir\,dr\over r^2+a^2}+d\phi\right)\wedge d\theta
  \\
 \qquad\qquad\qquad\qquad
 +{(a^2|k|\cos^2\theta -r^2)r\over h}\left({dr\over r^2+a^2}+{ir\over h}\sin^2\theta d\phi\right)\wedge d\psi\biggr]&(k<0).
 \end{cases}
 \label{dTheta2_final}
\end{align}
It is a good consistency check that these vanish for $m=0$, because
$\Theta_I$ vanishes for the original LM geometries.  Using
\eqref{EOM-Bianchi_Theta_duality}, we can compute $\psih$:
\begin{multline}
 \delta \psih=-{(|k|+1)\,|m|\,a^{|k|+2}\,b\over \sqrt{2Q_1Q_5}} \,F
\biggl[
 \sin\theta\cos\theta \left(\mp {irdr\over r^2+a^2}-d\phi\right)\wedge d\theta
 \\
 +\cos^2\theta\left(-{rdr\over r^2+a^2}\pm {ir^2\sin^2\theta\over h}d\phi\right)\wedge d\psi
 \biggr].
 \label{psih_v-dep_pert}
\end{multline}
Both $\Theta_1$ and  $\Theta_2$ give the same  $\psih$, as they should.

Finally, let us turn to the almost hyperk\"ahler structure 2-forms,
$J^{(A)}$.  To consider their fluctuation, we must first fix the
zeroth order expression.  The flat metric \eqref{ds42_flat} can be
rewritten in the Gibbons-Hawking form as follows:
\begin{align}
 ds_4^2=V^{-1}(d\chi+\xi)^2+Vds_3^2,\label{jhaq5Jul13}
\end{align}
where
\begin{align}
\begin{split}
  V&={1\over \rho},\qquad ds_3^2=d\rho^2+\rho^2(d\vartheta^2+\sin^2\vartheta\, d\varphi^2),\\
 \sqrt{r^2+a^2}&=2\sqrt{\rho}\,\cos{\vartheta\over 2},\qquad
 r=2\sqrt{\rho}\,\sin{\vartheta\over 2},\\
 \phi&={\chi\over 2}-\varphi,\qquad \psi={\chi\over 2},\qquad \xi=(1+\cos\vartheta)d\varphi.
\end{split}
\end{align}
As the zeroth order basis, let us take
\begin{align}
 J^{(A)}=e^1\wedge e^{A+1}-{1\over 2}\epsilon^{ABC}e^{B+1}\wedge e^{C+1},
 \label{ivbj5Jul13}
\end{align}
where $A,B,C=1,2,3$ and
\begin{align}
\begin{split}
 e^1&=V^{-\half}(d\chi+\xi),\qquad
 e^2=V^{\half}d(\rho\sin\vartheta \cos\varphi),\\
 e^3&=V^{\half}d(\rho\sin\vartheta \sin\varphi),\qquad
 e^4=V^{\half}d(\rho\cos\vartheta).
\end{split}
\end{align}
$e^2,e^3,e^4$ give the Cartesian coordinate basis of the base $\bbR^3$.
We could have instead taken the four Cartesian coordinate basis forms of
$\cB_4=\bbR^4$ as the zeroth order, but the above choice is
more in line with the circular profile function of the background LM
geometry.

With the above choice of $J^{(A)}$, the fluctuation $\delta J^{(A)}$ are
found to be
\begin{align}
 \delta J^{(1)}
 & =
 {(1+|k|)\, a^{1+|k|} \, b\,  Fe^{\pm i(\phi-\psi)}\over (r^2+a^2)^{3/2}}
 \notag\\
 &\quad \times
 \biggl[
 \pm i \left(\half \left[a^2+(a^2+2r^2)\cos(2\theta)\right]dr\wedge d\theta 
 -r(r^2+a^2)\sin\theta \cos\theta \,d\phi \wedge d\psi \right)
 \notag\\
 &\qquad -\cos\theta \sin\theta \, dr\wedge 
 \left[(r^2+a^2)d\phi -r^2 d\psi\right]
 +r(r^2+a^2)d\theta \wedge (\sin^2\theta\, d\phi -\cos^2\theta\, d\psi)
 \biggr],
 \notag
 \\
 \delta J^{(2)}
 & =
 {(1+|k|)\, a^{1+|k|} \, b\,  Fe^{\pm i(\phi-\psi)}\over (r^2+a^2)^{3/2}}
 \notag\\
 &\quad \times
 \biggl[
 \left(\half \left[a^2+(a^2+2r^2)\cos(2\theta)\right]dr\wedge d\theta 
 -r(r^2+a^2)\sin\theta \cos\theta \,d\phi \wedge d\psi\right) 
 \notag\\
 &\qquad \pm i\cos\theta \sin\theta \, dr\wedge 
 \left[(r^2+a^2)d\phi -r^2 d\psi\right]
 \mp i r(r^2+a^2)d\theta \wedge (\sin^2\theta \,d\phi -\cos^2\theta \,d\psi)
 \biggr],
 \notag
 \\
 \delta J^{(3)}
 &=
 (1+|k|)\, a^{1+|k|} \, b\,  F 
 \sin (2\theta )\left( {\pm ir\over r^2+a^2}dr\wedge d\theta -d\theta\wedge  d\phi\right).\label{dJ_final}
\end{align}
For details of the computation, see Appendix \ref{app:dJ(A)}\@.  One can
check that the above $\delta J^{(A)}$ correctly give $\delta \psih$
given in \eqref{psih_v-dep_pert} using the definition \eqref{psidef}.

\section{Future directions}
\label{sec:future}

In this paper, we perturbatively constructed supersymmetric
configurations of the D1-D5-P system as solutions of 6D supergravity at
the linear order.  An important characteristic of our solutions is that
they has $v$-dependent base space $\cB_4$.  This is a feature expected
of superstratum solutions \cite{Bena:2011dd} and we hope that our
solutions are useful for constructing general superstrata.

Our solutions have AdS asymptotics, because we used the solution
generating technique of \cite{Mathur:2003hj}.  It would be interesting
if our solutions can be generalized to flat asymptotics.  This is a
non-trivial problem, because adding ``1'' to the harmonic functions
$Z_{1,2}$ affect other equations in section \ref{sec:review_6d} and
finding $\Theta_{1,2},\omega,\cF$ that satisfy them is not an obvious
task.
Also, it is interesting to see how our solutions fit in the framework of
\cite{Niehoff:2013kia}, which discusses $v$- and $\chi$-dependent
fluctuations on top of $v$- and $\chi$-independent Gibbons-Hawking base.
Finally, our solutions are constructed as linear perturbations around
the maximally rotating Lunin-Mathur geometry.  It would be interesting
to see if this perturbative solution can be non-linearly completed to
finite deformations of the LM geometry \cite{Giusto:2013rxa}.  This will
make it easier to see the location of the brane sources in our
solutions, which should be useful for finding general smooth solutions
of the 6D system.

\section*{Acknowledgments}

I would like to thank Iosif Bena, Borun Chowdhury, Stefano Giusto, Samir
Mathur, Daniel Mayerson, Ben Niehoff, Rodolfo Russo, Yogesh Srivastava
and Nick Warner for valuable discussions.  In particular, I would like
to thank Stefano Giusto and Rodolfo Russo for useful comments on the
manuscript.  This work was supported in part by Grant-in-Aid for Young
Scientists (B) 24740159 from the Japan Society for the Promotion of
Science (JSPS)\@.

\appendix

\section{Convention}
\label{app:conv}

We define the following operators
\begin{align}
 D&\equiv\tilde d-\beta\wedge \partial_v,\\
 \dot{~}&\equiv\partial_v
 \equiv\cL_{\partial\over\partial v}
 =\iota_{\partial\over\partial v}^{} d+d\iota_{\partial\over\partial v}^{}
 ~.
 \label{Ddef}
\end{align}
The Hodge star is defined by
\begin{align}
 *_d\, (dx^{m_1}\wedge\cdots\wedge dx^{m_p})
 &={1\over (d-p)!}
 dx^{n_1}\wedge\cdots\wedge dx^{n_{d-p}}\,
\epsilon_{n_1\dots n_{d-p}}{}^{m_1\dots m_p}.
\label{Hodgestardef}
\end{align}
Our choice for the 6D $\epsilon$ tensor is \cite{Bena:2011dd}
\begin{align}
 \epsilon^{vu1234}= \epsilon^{ty1234}
 =+{1\over\sqrt{|g|}},\qquad
 \epsilon_{ty1234}=-\sqrt{|g|}.
\end{align}

\section{Lunin-Mathur geometry}
\label{app:LM_geom}

Here we summarize relations relevant for the Lunin-Mathur solutions
presented in \eqref{LM_in_GMR_form}.  

The periodicity of the profile functions, $L$, is related to the radius $R$
of the $S^1$ and the quantized D5 charge $N_5$ as
\begin{align}
 L={2\pi g_s \ap N_5\over R}.
 \label{L_def}
\end{align}
Given the profile function $F_m(w)$,
D1 charge is given by
\begin{align}
 Q_1={Q_5\over L}\int_0^L|\dot{F}|^2dw.
 \label{Q1_rel}
\end{align}
D1 charge $Q_1$ and D5 charge $Q_5$ are related to quantized charges
$N_1,N_5$ by
\begin{align}
 Q_1&=\gs\alpha'N_1,\qquad
 Q_5={\gs\alpha'^3 \over v_4}N_5,\qquad
 \label{Q_and_N}
\end{align}
where the coordinate volume of $T^4$ is $(2\pi)^4v_4$.

The 1-form $B$ can be found by solving the differential equation
$dB=*_4dA$ in \eqref{LM_in_GMR_form}.  The explicit solution is
\begin{align}
B=
 -{Q\epsilon_{ijkl}\over L}\int_0^L dw\int_0^1 dt\,
 {t\dot F_k F_l (y_i dx_j-y_j dx_i)\over |\vec y|^4},
\qquad y_i\equiv x_i-tF_i(w).\label{bicl5Jul13}
\end{align}
This can be derived as follows.  Let us rewrite the expression for $A$
in \eqref{LM_in_GMR_form} by decomposing the closed curve
$\vec{x}=\vec{F}(w)$ into sum of many closed curves, just like one does
in Stokes' theorem.
\begin{align}
 A
 &=
 -{Q_5\over L}
 \int_0^L dw \int_0^1 dt{\p\over\p t}\left[{t\dot{F}_i(w)dx_i\over |\vec x-t\vec F(w)|^2}\right]\notag\\
 &=
 -{Q_5\over L}
 \int_0^L dw \int_0^1 dt \left[
 {\dot{F}_i(w)dx_i\over |\vec x-t\vec F(w)|^2}
 +{2((\vec x-t \vec F)\cdot \vec F)\,t\dot{F}_i(w)dx_i \over |\vec x-t\vec F(w)|^2}
 \right]
\end{align}
This corresponds to decomposing the closed curve $\vec{x}=\vec{F}(w)$ as
a sum of many curves $\vec{x}=(t+dt)\vec{F}(w)$ and $\vec{x}=-t
\vec{F}(w)$.  The curves are along $w$, but we further want to divide
them by adding segments along $t$, so that now we have infinitesimal
curves along both $t$,$w$ directions.  This can be done by adding a
total derivative in $w$ (which integrates to zero upon $\int dw$) as
follows:
\begin{align}
 A
 &=
 -{Q_5\over L}
 \int_0^L dw \int_0^1 dt \left[
 {\dot{F}_i(w)dx_i\over |\vec x-t\vec F(w)|^2}
 +{2((\vec{x}-t\vec{F})\cdot \vec{F})\,t\dot{F}_i(w)dx_i \over |\vec x-t\vec F(w)|^2}
 -{\p\over \p w}\left( {F_i(w)dx_i\over |\vec x-t\vec F(w)|^2}\right)
 \right].
\end{align}
After some manipulation, this can be written as
\begin{align}
 A&=
 {2Q\over L}\int_0^L dw\int_0^1 dt\,
 {t\dot F_i F_j (y_i dx_j-y_j dx_i)\over |\vec y|^4},
\qquad y_i\equiv x_i-tF_i(w).
\end{align}
Now, if we have a 1-form
\begin{align}
 a=a_{ij}{x_i dx_j-x_j dx_i\over |\vec x|^4},
\end{align}
where $a_{ij}$ is constant and antisymmetric, then the 1-form $b$ that
satisfies
\begin{align}
 da=*_4 db
\end{align}
is given by
\begin{align}
 b=b_{ij}{x_i dx_j-x_j dx_i\over |\vec x|^4},\qquad
 b_{ij}=-{1\over 2}\epsilon_{ijkl}a_{kl}=-\tilde a_{ij}.
\end{align}
Therefore, \eqref{bicl5Jul13} is the solution to $dB=*_4dA$.

\section{Details of calculations}
\label{app:detail}

Here we describe some details of the computation in section 
\ref{sec:construction}.

\subsection{Fluctuation of Lunin-Mathur geometry}

We study fluctuations of the LM geometry corresponding to the
fluctuation $\delta F_m(w)$ of the profile function around the
background profile $F_m(w)$.  The change in the harmonic functions in
\eqref{LM_in_GMR_form} is given by
\begin{align}
\begin{split}
  \delta Z_1
 &={2Q_5\over L}\int_0^L dw
 \left[
 {((\vec{x}-\vec{F})\cdot \delta\vec{F})\dot{F}^2
 \over |\vec{x}-\vec{F}|^4}
 +{\dot{F}\cdot \delta\dot{\vec{F}}
 \over |\vec{x}-\vec{F}|^2}
 \right],
 \\
 \delta Z_2
 &={2Q_5\over L}\int_0^L dw
 {(\vec{x}-\vec{F})\cdot \delta\vec{F}\over |\vec{x}-\vec{F}|^4},
  \\
 \delta A_i
 &=-{Q_5\over L}\int_0^L dw
 \left[
 {2((\vec{x}-\vec{F})\cdot \delta\vec{F})\dot{F}_i
 \over |\vec{x}-\vec{F}|^4}
 +{\delta\dot{F}_i
 \over |\vec{x}-\vec{F}|^2}
 \right].
\end{split}
 \label{dZ_and_dA}
\end{align}
Also, from \eqref{bicl5Jul13},
The change in D1 charge $Q_1$ defined in \eqref{Q1_rel} is
\begin{align}
 \delta Q_1={2Q_5\over L}\int_0^L dw \,
 \vec{F}\cdot\delta\vec{F}.
\end{align}

For studying fluctuations around the maximally rotating LM solution
\eqref{MRLM_profile}, it is useful to define
\begin{align}
 I_n(k)&\equiv
 {1\over 2\pi}\int_0^{2\pi}
 {\cos(k\gamma)\,d\gamma\over (s^2+a^2+w^2-2as\cos\gamma)^n}
 = I_n(-k),
\label{I_n(k)_def}
\end{align}
for $k\in\bbZ$ and $n=1,2,\dots$.
Explicitly,
\begin{align}
 I_1(k)&={a^{|k|}\sin^{|k|}\theta\over h\,(r^2+a^2)^{|k|/2}},\label{I1}\\
 I_2(k)&=
 {\bigl[(|k|+1)r^2+((|k|-1)\cos^2\theta+2)a^2\bigr]
 a^{|k|}\sin^{|k|}\theta
 \over h^3\,(r^2+a^2)^{|k|/2}}\label{I2}.
\end{align}
We also define
\begin{align}
 X_\pm&\equiv as[I_2(k+2)\pm I_2(k)]+a^2[\mp I_2(k-1)-I_2(k+1)]+(k+1)I_1(k+1).
 \label{Xpm_def}
\end{align}
More explicitly,
\begin{align}
 X_+&=\begin{cases}
       {\left[|k|\left((1-2\cos^2\theta)r^2-a^2\cos^2\theta\right)h
 +2(r^2+a^2)(r^2-a^2\cos^2\theta)\sin^2\theta\right]a^{|k|+1}\sin^{|k|-1}\theta
 \over (r^2+a^2)^{|k|+1\over 2}h^3
} & (k\neq 0),\\[2ex]
       {2\sqrt{r^2+a^2}(r^2-a^2\cos^2\theta)a\sin\theta\over h^3}
       & (k=0).
\end{cases}
 \\
X_-&={ka^{|k|+1}\sin^{|k|-1}\theta\over (r^2+a^2)^{|k|+1\over 2}h}.
\end{align}

\subsection{The seed solution and spectral flow}
\label{app:seed_spectral_flow}

In section \ref{ss:seed_and_spectral_flow}, we considered the
fluctuation \eqref{bjga5Jul13} around the maximally rotating LM geometry
and computed the change in the GMR data.  The change in $Z_1,Z_2,A,B$ is
straightforward to compute using the formulas
\eqref{dZ_and_dA} and \eqref{bicl5Jul13}.  
The change in the RR 2-form \eqref{C2_LM}, $\delta C_2$, has 
contributions $\delta C_{2,\rm elec}$ and $\delta C_{2,\rm mag}$:
\begin{align}
 \delta C_{2,\rm elec}&=-{\delta Z_1\over Z_1^2}(dt-A)\wedge (dy+B)
 +Z_1^{-1}[-\delta A\wedge (dy+B)+(dt-A)\wedge \delta{B}],\\
 \delta C_{2,\rm mag}&=\delta \cC_2,\qquad
 d\delta \cC_2=*_4dZ_1.
\end{align}
If we carry out the spectral flow \eqref{spectralflow} followed by the
coordinate transformation \eqref{bjoy5Jul13}, we have an additional
contribution:
\begin{align}
 (\delta C_{2,\rm diff})_{\mu\nu}
= (\cL_\xi C_{2})_{\mu\nu}
=
 \xi^\rho \partial_\rho C_{\mu\nu}
 +\partial_\mu \xi^\rho C_{\rho\nu}
 +\partial_\nu \xi^\rho C_{\mu\rho},
\end{align}
where $\cL_\xi$ is the Lie derivative.
The total change in $C_2$ is given by
\begin{align}
 \delta C_2=\delta C_{2,\rm elec}+\delta C_{2,\rm mag}+\delta C_{2,\rm diff}.
\end{align}
In order to find $\delta C_2$ in the canonical form of
\cite{Deger:1998nm}, it is easier to first compute $\delta F_3=d\delta
C_2$, because then we do not have to know $\delta C_{2,\rm mag}$ but
only its exterior derivative $d\delta C_{2,\rm mag}=*_4dZ_1$.  After
some tedious computation, we find, for $k>0$,
\begin{align}
\delta F_3&={2 (k+1) a^{k-1} b e^{i k (t/R+\phit)} \sin ^k\theta  
 \over {Q_1 (r^2+a^2)^{k/2}}}
 \Biggl[-(k-2) r\, dt\wedge dr\notag\\
 &\qquad\qquad\qquad\qquad
 +k r^2 \left(dt-\frac{i a \sqrt{Q_1 Q_5}}{r(r^2+a^2)}dr\right)
 \wedge (\cot \theta  d\theta +i d\phit)\Biggr]
 \wedge dy\label{ipxn5Jul13}
\end{align}
while, for $k=-l<0$,
\begin{align}
\delta F_3
 &=
{2 a^{l-1} b e^{-i l (t/R+\phit)} \sin^l\theta 
 \over Q_1 (r^2+a^2)^{l/2}}
   \Biggl[(l-2) r \,dt\wedge dy\wedge dr
 +l (l+2) Q_1 Q_5 \sin\theta  \cos \theta\,  d\theta\wedge d\phit\wedge d\psit
 \notag\\
 &\qquad\qquad
 -l r^2 \left(dt+\frac{i\sqrt{Q_1 Q_5}\, a}{r(r^2+a^2)}dr\right)
 \wedge dy\wedge (\cot\theta\,  d\theta -i d\phit)
 \notag\\
 &\qquad\qquad
 -l^2   a \sqrt{Q_1 Q_5} \left(dt-\frac{i\sqrt{Q_1 Q_5}\, r}{a
   \left(r^2+a^2\right)}dr\right)
 \wedge \left(\cot \theta\,  d\theta -i \cos ^2\theta \, d\phit\right)
 \wedge d\psit\Biggr].\label{ipzk5Jul13}
\end{align}
Note that the expression for $k=-l<0$ is not simply obtained from the one
for $k>0$ by replacing $k\to l$.

The 2-form potential $\delta C_2$ that gives the above $\delta F_3$ is
obtained as follows.  First, from \cite{Deger:1998nm}, the AdS$_3$ part
of the 2-form can be written as
\begin{align}
 C_{\mu\nu}=(\epsilon^{AdS_3})_{\mu\nu}{}^\lambda\, X_\lambda\, \Yh,
\end{align}
where $\mu,\nu,\lambda$ are AdS$_3$ indices and $\epsilon^{AdS_3}$ is
the volume form for AdS$_3$ with the metric 
\eqref{AdS3_metric}. $X_\lambda$ are functions in AdS$_3$ while $\Yh$ is
a harmonic function in $S^3$.  On the other hand, the $S^3$ part can be
written as
\begin{align}
 C_{ab}=(\epsilon^{S^3})_{ab}{}^c\, U\, \partial_c \Yh,
\end{align}
where $a,b,c$ are $S^3$ indices, $\epsilon^{S^3}$ is the volume form for
unit $S^3$ with the metric \eqref{S3_metric}, and $U$ is a function in
AdS$_3$.  In general, there can be also mixing terms, $C_{\mu a}$, but
that turns out unnecessary in the present case. So, after a bit of
redefinitions, our ansatz for the 2-form is
\begin{align}
 \delta C_2
 &=\Bh
 \left[
  X_t \,{r\over r^2+a^2}dy\wedge dr
 + X_y {dr\wedge dt\over r}
 +X_r {r(r^2+a^2)\over Q_1Q_5}dt\wedge dy
 \right]\Yh\notag\\
 &\qquad +
 \Bh U
 \left[
 \sin\theta\cos\theta (\partial_\theta \Yh)d\phit\wedge d\psit
 +{\cos\theta\over\sin\theta} (\partial_\phit \Yh)d\psit\wedge d\theta
 +{\sin\theta\over \cos\theta} (\partial_\psi \Yh)d\theta\wedge d\phit
 \right],
 \end{align}
where $\Bh,\Yh$ are defined in \eqref{def_B,Y}.  By requiring that this
reproduce the 3-form $\delta F_3$ in \eqref{ipxn5Jul13} and
\eqref{ipzk5Jul13}, we get the following simple result:
\begin{align}
 k>0&:\quad 
 X_t=-\frac{2 i (k+1) a^{k-1} b}{\omega},\quad
 X_y=0,\quad
 X_r=-\frac{2 (k+1) a^{k-1}b Q_5 r }{r^2+a^2},\quad
 U= 0,
 \notag\\
 k=-l<0&:\quad 
 X_t=\frac{2 i a^{l-1} b}{\omega },\quad
 X_y=0,\quad
 X_r=-\frac{2 a^{l-1} b Q_5 r }{r^2+a^2},\quad
 U=-2 a^{l-1} b Q_5 .
  \label{dC2nonlong}
\end{align}
Or, more explicitly,
\begin{align}
\delta C_2=
\begin{cases}
 \displaystyle -{2 (k+1) a^{k-1} b \over Q_1 \omega}
 \Bh\Yh
\left[r^2 \omega\,  dt\wedge dy+i Q_1 \frac{r\, dy\wedge dr}{r^2+a^2}\right]& (k>0),
 \\[3ex]
 \displaystyle  -{2 a^{l-1} b \over Q_1 \omega}\Bh\Yh\biggl[r^2 \omega\,  dt\wedge dy
 -i Q_1 \frac{r\, dy\wedge dr}{r^2+a^2}
 \\[2ex]
 \displaystyle \qquad\qquad\qquad
 -ilQ_1Q_5\omega \cot\theta
 (d\theta -i\sin\theta\cos\theta d\phit)\wedge d\psit\biggr]
 &(k=-l<0).
\end{cases}
\end{align}
This is what we used in \eqref{delta_C_2}.

\subsection{Computing $\delta J^{(A)}$}
\label{app:dJ(A)}

As explained in the main text, as the zeroth order solution, we used the
hyperk\"ahler structure 2-forms $J^{(A)}$ defined through the vierbein
$e^I=e^I{}_i dx^i$, $I=1,2,3,4$, as \eqref{ivbj5Jul13}.  Note that
$J^{(A)}$ are genuinely hyperk\"ahler, not almost hyperk\"ahler, and
therefore closed.  Also, note that $e^I$ are orthonormal in the sense
\begin{align}
 g_{4}^{ij} e^I{}_i e^J{}_j =\delta^{IJ},\label{jhpe5Jul13}
\end{align}
where $g_{4}^{ij}$ is the inverse of the base metric $g_{4\,ij}$ 
defined in \eqref{jhaq5Jul13}.

Let us assume that the corrected 2-forms $J^{(A)}+\delta J^{(A)}$ are
still constructed from the corrected vierbein $e^I+\delta e^I$ by
\eqref{ivbj5Jul13}.  Namely,
\begin{align}
 \delta J^{(A)}=\delta e^1\wedge e^{A+1}+ e^1\wedge \delta e^{A+1}
 -\half\epsilon^{ABC}(\delta e^B\wedge e^C+e^B\wedge \delta e^C).\label{dJ_ito_e,de}
\end{align}
Let us expand $\delta e^{I}$ as $\delta e^I=\delta e^I{}_i dx^i$ and
raise and lower indices using the zeroth order quantities $e^{I}{}_j$,
$g_{4\, ij}$, and $g_4^{ij}$. If we require that $e^I+\delta e^I$ be
orthonormal with respect to the corrected metric $g_4+\delta g_4$, then
\eqref{jhpe5Jul13} implies that
\begin{align}
 \delta e_{ij}+\delta e_{ji}=\delta g_{4\, ij}.
\end{align}
Therefore, we can write $\delta e^I$ in terms of the 6 independent
variables $\delta e_{i<j}$ as
\begin{align}
 \delta e^I=\sum_{i=1}^4 \left[\half e^{Ii}\delta g_{4\,ii}+
 \sum_{1\le j<i}e^{Ij}\delta e_{ji}+\sum_{i<j\le 4}e^{Ij}(\delta g_{4\,ij}-\delta e_{ij})\right]dx^i.\label{jror5Jul13}
\end{align}
With this construction, the conditions \eqref{J_ASDty} and
\eqref{J_quat} on $J^{(A)}+\delta J^{(A)}$ are automatically satisfied.
However,  they will not be closed any more.

In the present case, all fields \eqref{dGMR_data_final},
\eqref{dTheta1_final}, and \eqref{dTheta2_final} depend on $v$ through
$F$ defined in \eqref{def_F}.  So, let us assume that $\delta e,\delta
J^{(A)}$ are also proportional to $F$ and therefore
\begin{align}
 \delta \dot J^{(A)}=i\sqrt{2\over Q_1Q_5}\,am\, \delta J^{(A)}.
\end{align}
In this case, $\psih$ in \eqref{psidef} is given by
\begin{align}
 \psih
 &=i\sqrt{2\over Q_1Q_5}\, am\cdot{1\over 16}\epsilon^{ABJ}J^{(A)ij}\delta J^{(B)}_{ij}J^{(C)}\notag\\
 &={i\over \sqrt{8Q_1Q_5}am}(1-*_4)M,\qquad M=e^I\wedge \delta e_I.
 \label{jrqo5Jul13}
\end{align}
If we plug the explicit expression \eqref{jror5Jul13} into
\eqref{jrqo5Jul13} and require that it be equal to
\eqref{psih_v-dep_pert}, it turns out that we can eliminate 3 out of 6
independent parameters $\delta e_{i<j}$.  For example, we can take
$\delta e_{12},\delta e_{13}, \delta e_{14}$ as independent variables.

One can show that the differential condition \eqref{J_diff_cond}, which
reads
\begin{align}
 \tilde{d}\delta J&=\partial_v(\beta\wedge \delta J+\delta B\wedge J)
 \qquad\\
 &=i\sqrt{2\over Q_1Q_5}ma(\beta\wedge \delta J+\delta B\wedge J),
\end{align}
is identically satisfied, whatever the values of $\delta e_{12},\delta
e_{13}, \delta e_{14}$ are.  If we compute $\delta J^{(A)}$ using
\eqref{dJ_ito_e,de}, we obtain \eqref{dJ_final}, independent of 
$\delta e_{12},\delta
e_{13}, \delta e_{14}$.

\appendix


\end{document}